# Calculation of the transport critical current density of c-axis textured 122 iron-based superconductors

Lei Wang[*], Jinghui Xu, and Duoduo Ba

Department of Power and Electrical Engineering, Northwest A&F University, Yangling, Shaanxi 712100, China

Xianping Zhang, Zhaoshun Gao, Dongliang Wang, and Yanwei Ma

Key Laboratory of Applied Superconductivity, Institute of Electrical Engineering, Chinese Academy of Sciences, P. O. Box 2703, Beijing 100190, China

**Abstract**

The c-axis textured $Sr_{1-x}K_xFe_2As_2$ tapes produced by cold rolling and post-annealing, could carry a high super-current over $2\times10^4$ A/cm$^2$. However, the magnitude is far from its maximum, because of the current obstacles associated with various defects in the material. To predict the maximal transport critical current density, we modeled the current paths in a c-axis textured polycrystal as a three-dimensional flow network, and calculated the maximum flow with the Ford-Fulkerson algorithm. It indicates that a much higher super-current of about $2\times10^5$ A/cm$^2$ could be achieved in an ideal c-axis textured K-doped 122 polycrystal. The dependences of transport $J_c$ on density, content of invalid boundary and grain size and shape were also studied. The results imply that, over 30% of the grain boundaries in the reported c-axis textured $Sr_{1-x}K_xFe_2As_2$ tapes may act as current obstacles, and the large ratio of width to thickness was expected to be the most favorable grain shape for high transport $J_c$ in c-axis textured 122 superconducting tapes.

[*] Author to whom correspondence should be addressed; E-mail: leiwang@nwsuaf.edu.cn

**Introduction**

Since the discovery of superconductivity in LaFeAsO$_{1-x}$F$_x$ in 2008,[1] the current carrying capability of polycrystalline iron-pnictide superconductors has been a topic of great technological interest due to their possible industrial applications.[2-10] Some previous works on iron-pnictide bicrystal grain boundary junctions indicated that the critical current density through the grain boundary ($J_c^{GB}$) decay less rapidly with increasing misorientation angle than that for YBa$_2$Cu$_3$O$_{7-\delta}$.[11, 12] Given this encouragement, researchers attempted to fabricate iron-pnictide conductors with a high transport critical current density ($J_c$).[13-18] Recently, great progresses have been achieved in this attempt, and over one hundred amperes of current ($J_c > 10^4$ A/cm$^2$) has been obtained in A$_{1-x}$K$_x$Fe$_2$As$_2$ (A=Ba, Sr)[19] superconducting wires and tapes,[20] which were fabricated by revised powder-in-tube methods. In particular, the c-axis texture, produced by a cold rolling process with a subsequent annealing, could significantly depress the weak-link effect of large-angle grain boundary and improve current-carrying capability, appearing to be a much promising and practical approach.[21] By this approach, a high transport $J_c$ of $2.5 \times 10^4$ A/cm$^2$ was achieved in Sn-added Sr$_{1-x}$K$_x$Fe$_2$As$_2$ tapes.[22] As the super-current in the textured iron pnictide was also obstructed by defects, such as impurity phases, cracks, pores and amorphous layers,[3, 6, 23] the transport $J_c$ could be further enhanced by optimizing the fabrication and annealing procedures. Whereas, a key question is, how high the transport $J_c$ can be achieved, if all those defects in the c-axis textured A$_{1-x}$K$_x$Fe$_2$As$_2$ (A=Ba, Sr) were eliminated.

In this paper, we take the current paths in a c-axis textured polycrystalline as a three-dimensional flow network, and calculate the transport $J_c$ with the Ford-Fulkerson algorithm.[24, 25] Density, invalid boundary and grain size and shape were also considered in the model.

**Model and calculation**

For simplicity, a current model for one-dimensional c-axis textured polycrystalline superconductor was considered at first, as shown in figure 1. There are n+1 grains with n overlapping connections. If the grains were weakly linked, that is, the current carrying capabilities of all the overlapping connections ($J_i$, where i=1, 2, 3, ... , n) were much weaker than that of grains, the maximal current that can flow through this one-dimensional polycrystalline superconductor ($J_c$) is just the minimum of $J_i$. Thus, for a one-dimensional weak-linked polycrystalline superconductor, the transport $J_c$ is determined not by the grains but the weak-linked grain boundaries. Mathematically, the transport $J_c$ is a function of $J_i$ (i=1, 2, 3, ... , n). Since all grains are c-axis textured, $J_i$ is actually the critical current of a twist grain boundary, for which the misorientation angle could be any value from 0 to 360º.

For real c-axis textured polycrystalline bulk, the current paths constitute a three-dimensional network. Currents can go any way possible through the bulk. If we could find all possible ways (this job can easily be done by computer), the critical current of the polycrystalline bulk can be obtained by summing the $J_c$ for each way. The only restriction is that the sum of all currents that flow through a twist grain boundary can not exceed its current carrying capability $J_i$. Mathematically, the network of current paths is called flow network, and the solving method described above is called Ford-Fulkerson algorithm. [24, 25]

Here we use a simple three-dimensional current model, as shown in fig. 2a, to predict the transport properties of c-axis textured 122 iron pnictides. All the grains are assumed oriented with their common c-axis along a given direction, while the orientation of the a and b axes of the grains is randomly normal to that given direction, which closely resembles the experimentally observed microstructure. For simplicity, we assumed that each grain of a upper layer overlaps four grains of the lower, and the electrical contact between the adjacent grains of the same layer was neglected.

The current paths in the model constitute a three-dimensional flow network, as shown in fig. 2b. The current paths in grains were represented by nodes and the paths across twist boundaries were represented by arcs. Each arc has a capacity, and the amount of flow on an arc can not exceed the capacity of the arc. A flow must also satisfy the restriction that the amount of flow into a node equals the amount out of it. We also assumed that all arcs in the network have an orientation (represented by the arrows in the network), along which arcs can only receive flow.[25] We solve the maximum flow problem using a Matlab program with the Ford-Fulkerson algorithm.

The only critical issue left is the determination of current carrying capability of various twist grain boundary. The weak-link effect of $BaFe_{2-x}Co_xAs_2$ has been studied experimentally in bicrystals.[11, 12] It indicated that the critical current density through [001] tilt grain boundary ($J_c^{GB}$) drop exponentially with increasing misorientation angle, but less rapidly than that for $YBa_2Cu_3O_{7-\delta}$. We analyzed these results and found that the observed $J_c^{GB}$ seems consistent with the phenomenological equation:

$$J_c^{GB} = [\frac{2000000}{\mu_0 H + 0.4}\exp(-\frac{\theta}{3.5}) + (\frac{38000}{\mu_0 H + 0.4} + 3800)\exp(-\frac{\theta}{30 + \mu_0 H})](1-\frac{T}{T_c})^2$$

where $J_c^{GB}$ is the transport $J_c$ across the twist boundary in A/cm$^2$, $\mu_0 H$ is the applied field in Tesla, $\theta$ is the miorientation angle (0º ≤ $\theta$ ≤ 45º), $T$ is the temperature, and $T_c$ is the critical transition temperature. For 45º ≤ $\theta$ ≤ 90º, it is reasonable to replace the $\theta$ in the equitation by 90-$\theta$, because of the symmetry of the iron pnictide lattice. Figure 3 shows the calculated $J_c^{GB}$ value from the given equation (colored lines) and the experimental data[12] (red circles for 0.2 T, green triangles for 1 T and blue diamonds for 9 T) for $BaFe_{2-x}Co_xAs_2$ with a $T_c$ of 21 K. One can see that this equitation is well consistent with the experimental results. To calculate the transport $J_c$ of c-axis textured $A_{1-x}K_xFe_2As_2$ (A=Ba, Sr), we assumed here that the $J_c^{GB}$ for $A_{1-x}K_xFe_2As_2$ (A=Ba, Sr) follow the given equation as well, and that the miorientation angle $\theta$ in the equation can be a tilt angle for a [001] tilt grain boundary or a twist angle for [001] twist grain boundaries.[26]

It should be noted that the pinning effects of dislocations at the grain boundary has been implicitly contained in the above equation, while artificial pinning centers and related pinning effects are not considered in this model.

**Results and discussion**

First we calculated the transport $J_c$ of a small ideal c-axis texted $A_{1-x}K_xFe_2As_2$ (A=Ba, Sr, $T_c$ =38 K) polycrystal (with a size of 40×200×400 μm$^3$), which consists of 16000 grains, leaving out all the defects like porosity, impurity phases, or any other current obstacles. The grain size supposed here is 2×10×10 μm$^3$, which is a common size observed in 122 superconducting tapes and wires.[21, 22] The transport $J_c$ versus magnetic field in the temperature range of 4 - 30 K was illustrated in fig. 4. One can see that a highest $J_c$ of almost 1×10$^5$ A/cm$^2$ can be achieved at 4 K close to self field, four times as high as the reported value for the c-axis textured $Sr_{1-x}K_xFe_2As_2$ superconducting tapes.[22] As the field increases to 5 T, the transport $J_c$ drops slowly to right over 1×10$^4$ A/cm$^2$. For further field enhancement, the transport $J_c$ almost keeps constant, maintaining about 7×10$^3$ A/cm$^2$ at a field as high as 20 T. Even at a temperature of 20 K, the calculated $J_c$ turn out be over 2×10$^4$ A/cm$^2$ close to self field. It should be noted that, although such a high transport $J_c$ has not been observed experimentally in c-axis textured 122 iron-pnictide superconductors yet, due to the presence of various defects in the material, the field dependence of the calculated $J_c$ is in good agreement with that of 122 superconducting tapes, for which the observed transport $J_c$ at 10 T is about one order of magnitude lower than that at self field.[21, 22] This result also confirms that our model and calculation are appropriate for the c-axis textured 122 iron pnictide.

For real polycrystalline superconductors, pores always exist, making the relative density $\rho_r$, which is the ratio of the density of a polycrystal to that of a corresponding single crystal, below 1. In this model, it was assumed that pores only exist between adjacent grains of the same layer. On this assumption, the transport $J_c$'s at 4 K and 20 K were calculated with respect to the relative density (from 0.3 to 1), as presented in fig. 5a and 5b respectively. The magnitude of $J_c$ decreases slowly with relative density decreasing, and a half of the maximum could be obtained at $\rho_r$=0.6-0.65. As relative density decreases down to 0.3, the transport $J_c$ remains only about three percent of its maximal value. Supposing a relative density of ~0.7 had been achieved for the Sn-added $Sr_{1-x}K_xFe_2As_2$ superconducting tapes,[22] the transport $J_c$ of ~ 2.5×10$^4$ A/cm$^2$ is less than a half of its calculated maximum (~ 6×10$^4$ A/cm$^2$), indicating that much work on eliminating current obstacles needs to be done. On the other hand, increasing density itself could be an important approach to obtain a high transport $J_c$. Actually, since the pores in this model only decrease the area of the twist boundaries, the variation of the calculated transport $J_c$ on relative density $\rho_r$ is given by,

$$J_c \propto \left(2 - \frac{1}{\sqrt{\rho_r}}\right)^2 \qquad (0.25<\rho_r<1)$$

In this model, the extrinsic current obstacles, such as cracks, impurity phases, and amorphous layers, at the grain boundaries were also considered.[3, 6, 23] The grain boundaries with such defects do not contribute to super-current carrying capability of the polycrystalline, and could be called invalid boundaries. The presence of the invalid boundaries can be modeled as a random cutting off of the arcs in the network, which consequently reduce the maximum flow, i.e. transport $J_c$. The decline in transport $J_c$ caused by invalid boundaries were shown in fig. 6a and 6b. As one can see, with a 0.25 content of invalid boundary, the transport $J_c$ is about one half of its maximum value. As the content reaching 0.4, only one forth of its maximum value remains. Recall the magnitude of transport $J_c$ reported for c-axis textured 122 superconducting tapes,[22] over 30% of the grain boundaries in the samples may be invalid. It is a hard, but critical work to eliminate the current obstruction defects, especially the amorphous layers, in the samples. The transport $J_c$ vanishes ($J_c<10^2$ A/cm$^2$) with about 0.65-0.7 invalid boundaries, or in other words, the transport $J_c$ appears with 0.3-0.35 valid boundaries, which is in agreement with the percolation threshold on three-dimensional lattices.[27]

The dependence of transport $J_c$ on grain size, in the range from 5 to 40 μm with a fixed ratio of width to thickness, was studied in this model, as shown in fig. 7a. It seems that the transport $J_c$ almost keep constant with various grain sizes (The decrease in the dimension of network makes the maximum flow a little scattering). It should be noted that, for this model, we supposed that all grains had been well textured, but for superconducting wire and tape fabrication, grain size may have a large effect on the texture deformation in the cold rolling process, and can affect the transport $J_c$ eventually. The dependence of transport $J_c$ on the ratio of width to thickness was also studied, and the results were shown in fig. 7b. Here the thickness was fixed to be 2 μm, with the width of grain ranging from 5 to 40 μm. More precisely, a restriction that amount of super-current through a twist boundary can not exceed the capacity of a grain was imposed here. Obviously, the transport $J_c$ increases with the ratio of width to thickness increasing. With a ratio of 20 (Width = 40 μm), a remarkable transport $J_c$ of $2\times10^5$ A/cm$^2$ can be achieved, twice as large as that for the ratio of 5 and four times as large as that for the ratio of 2.5. At a high field of 10 T, a transport $J_c$ of $\sim3\times10^4$ A/cm$^2$ can be obtained. The results suggest that the large ratio of width to thickness were expected to be the most favorable grain geometry for high transport $J_c$ in c-axis textured 122 superconducting tapes.

**Conclusions**

Our results indicate that a transport $J_c$ as high as $\sim2\times10^5$ A/cm$^2$ could be carried by an ideal c-axis textured K-doped 122 iron-pnictide superconductor, and a large value of $\sim3\times10^4$ A/cm$^2$ would remain at a high magnetic field of 10 T. The transport property is dependent on density as well as contents of invalid boundaries, and over 30% of the grain boundaries in the reported superconducting wires and tapes may act as current obstacles. It was also pointed out that refining the shape of grains is also a promising approach to obtain high transport $J_c$ in the c-axis textured 122 iron-pnictide superconductor.

**Acknowledgements**

This work is partially supported by National Science Foundation of China (Grant No. 51202200), the National '973' Program (Grant No. 2011CBA00105), National Science Foundation of China (Grant No. 51025726).

**Captions**

Figure 1 A current model for one-dimensional c-axis textured polycrystalline superconductor.

Figure 2 (a) A three-dimensional model. Each brick represents an iron-pnictide grain, with its c-axis along a given direction, and the a and b axes randomly normal to the direction. The electrical contact between the adjacent grains of the same layer was neglected.

(b) The current paths in the brick-wall model constitute a flow network, with the current paths in grains represented by nodes and that across twist boundaries represented by arcs.

Figure 3 The transport property of twist boundary with various misorientation angles. The lines show the calculated results from the given equitation, and the symbols (red circles for 0.2 T, green triangles for 1 T and blue diamonds for 9 T) show the measurements of $J_{cGB}$ for $BaFe_{2-x}Co_xAs_2$ bicrystal junctions.

Figure 4 The calculated transport Jc of a small (40×200×400 μm3) ideal c-axis texuted $A_{1-x}K_xFe_2As_2$ (A=Ba, Sr, Tc =38 K) polycrystal, with a grain size of 2×10×10 μm3.

Figure 5 The dependence of transport Jc on relative density (from 0.3 to 1) at 4 K (a) and 20 K (b), with the assumption that pores only exist between adjacent grains of the same layer.

Figure 6 The dependence of transport Jc on the content of invalid boundary at 4 K (a) and 20 K (b).

Figure 7 The variations of transport Jc originate from the grain size and shape effects (a) with a fixed width/thickness ratio and (b) a fixed thickness.

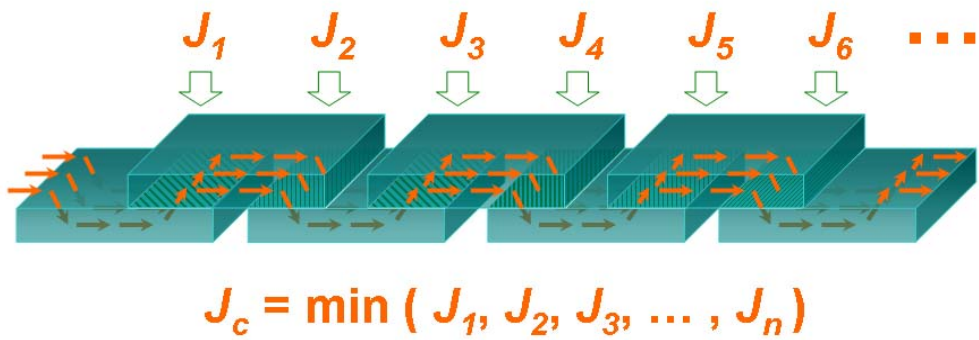

Figure 1 Wang et al.

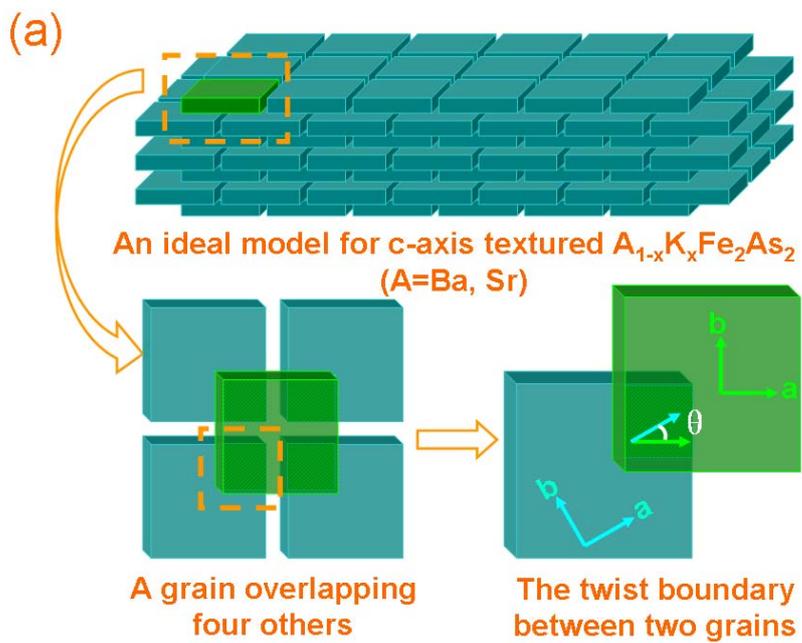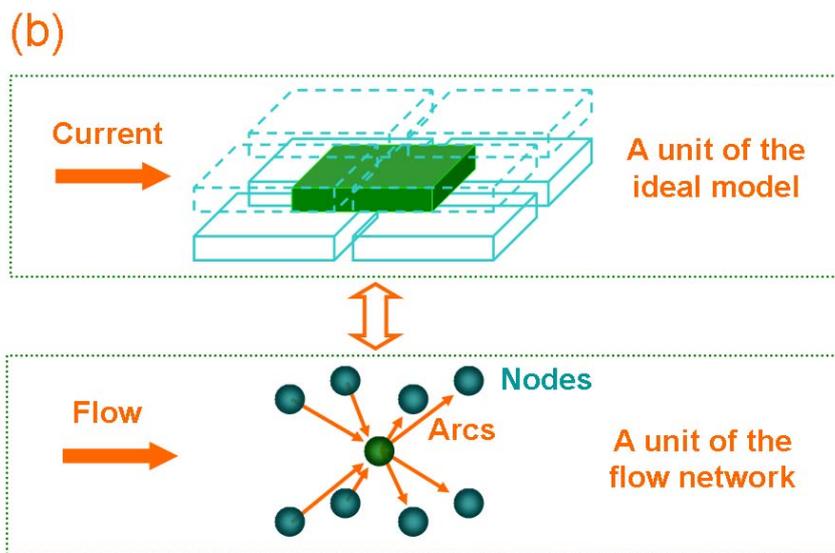

Figure 2 Wang et al.

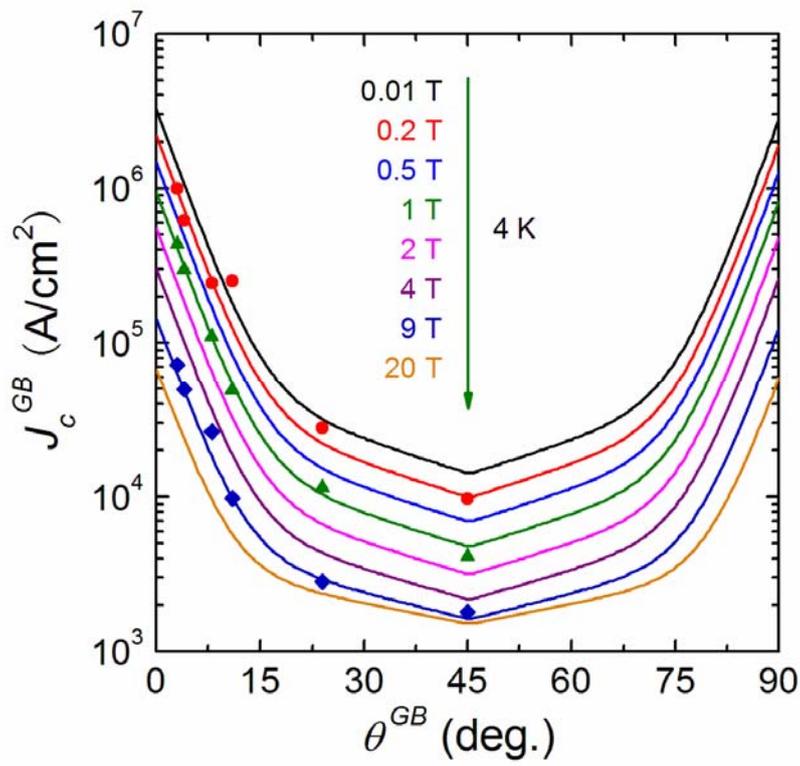

Figure 3 Wang et al.

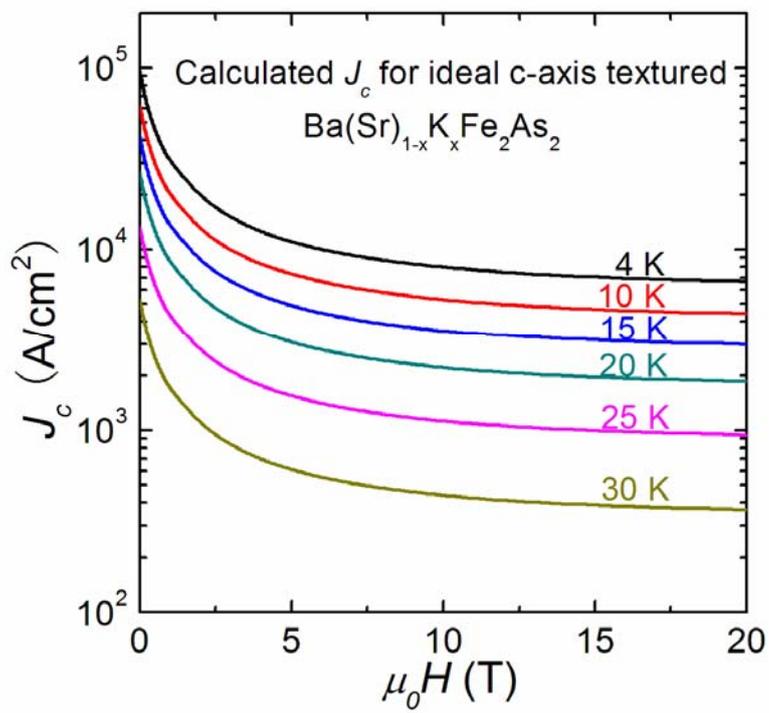

Figure 4 Wang et al.

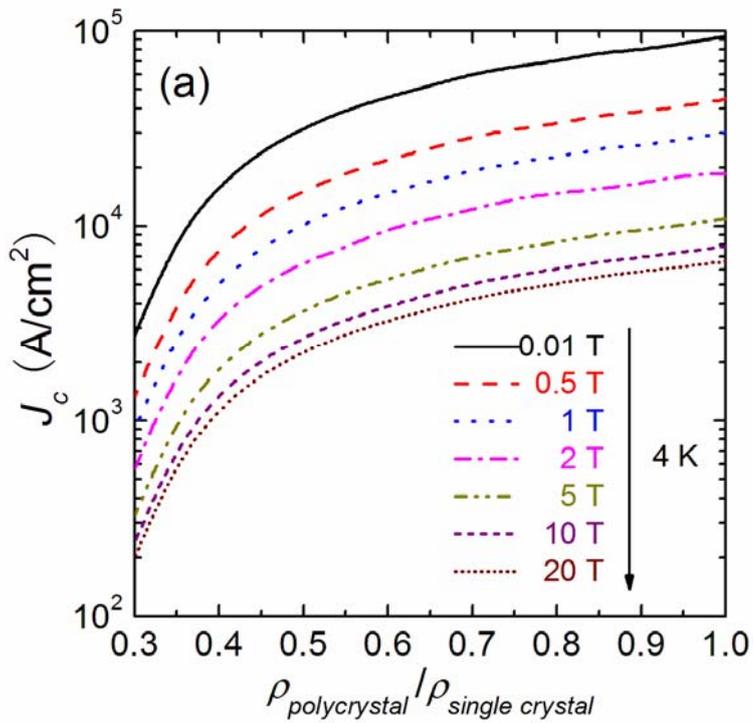

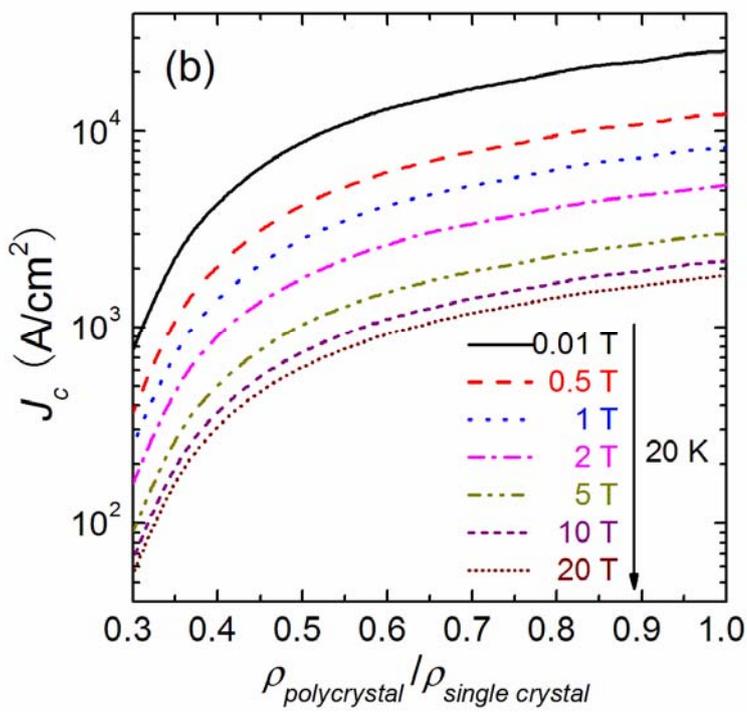

Figure 5 Wang et al.

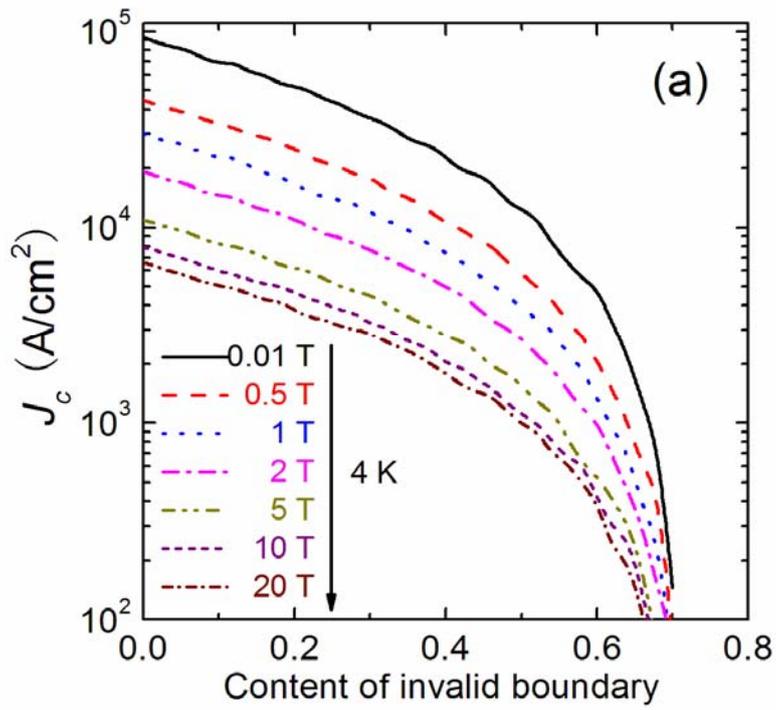
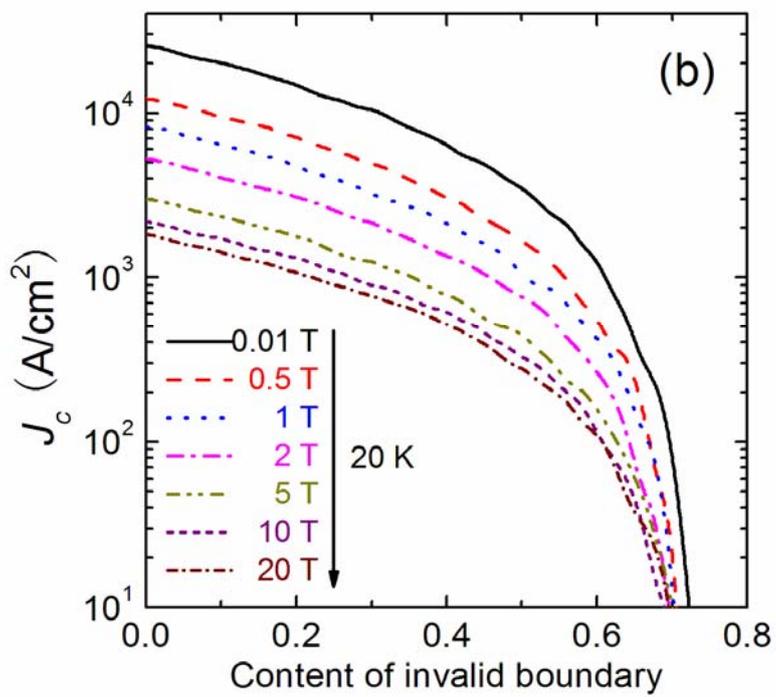

Figure 6 Wang et al.

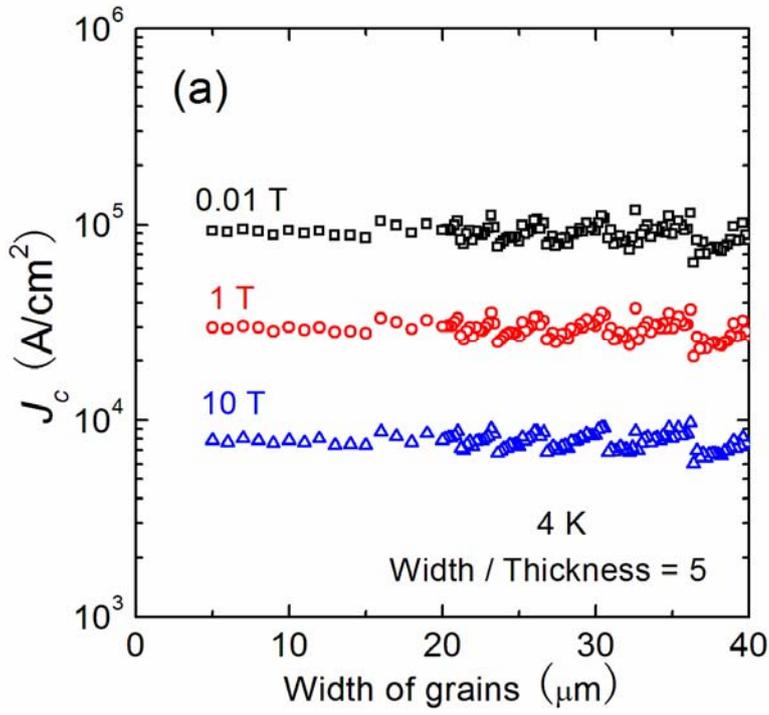

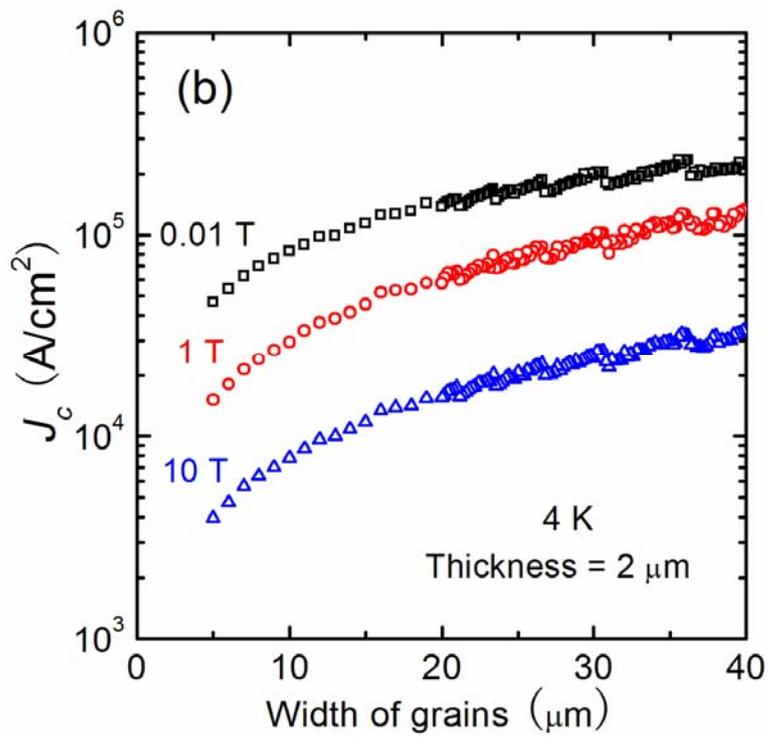

Figure 7 Wang et al.